\documentclass[reprint, amsmath, amssymb, aps]{revtex4}

\usepackage{graphicx}
\usepackage{dcolumn}
\usepackage{bm}
\usepackage[colorlinks,linkcolor=blue,anchorcolor=blue,citecolor=blue]{hyperref}
 
\begin{document}
\title{Microstates and statistical entropy of observed 4D black holes}
\author{Cao H. Nam$^{1,2}$}
\email{caohoangnam@duytan.edu.vn}  
\affiliation{$^1$Institute of Theoretical and Applied Research, Duy Tan University, Hanoi 100000, Vietnam\\
$^2$School of Engineering and Technology, Duy Tan University, Da Nang 550000, Vietnam}
\date{\today}

\begin{abstract} 
We aim to provide a microscopic explanation of observed 4D black holes based on the compactification of 5D Einstein gravity plus a positive cosmological constant on a circle. The framework of the dimensional reduction in this work allows us to compute the statistical entropy of general 4D black holes independent of the symmetries of the black hole solution, such as the spherical symmetry, and going beyond the class of special black holes that are supersymmetric and (near-)extremal as well as have exotic charges. The statistical entropy of 4D black holes includes the Bekenstein-Hawking area term at leading order and sub-leading exponential corrections. We find a new exponential correction that is more meaningful than those previously found in the literature.
\end{abstract}

\maketitle

\section{Introduction}

The pioneering studies of Bekenstein and Hawking showed that black holes behave as thermodynamic objects and have an entropy proportional to the area of their event horizon \cite{Bekenstein1973,Bardeen1973,Hawking1975}. But the nature of microscopic configurations (microstates) counted for the black hole entropy has not yet been totally understood, and a huge literature has been devoted to solving this problem. String theory can offer a statistical interpretation for the entropy of (near-)extremal and supersymmetric black holes \cite{Strominger1996,Horowitz1996,Maldacena1996,Johnson1996,Dabholkar2005,Boido2023}, however, these special black holes do not fall into the class of observed black holes. In loop quantum gravity, the calculation of the statistical entropy of black holes is performed under certain assumptions that are relevant to the introduction of the isolated horizon and the choice of the Immirzi parameter \cite{Ashtekar1998}. There have also been proposals that explain the statistical origin of the asymptotically flat non-rotating (charged) black hole entropy based on microstate geometries, which are smooth horizonless solutions \cite{Ibrah2021,Balasubramanian2023} and quasi-normal modes \cite{Cadoni2021}. But, it remains unclear how to provide a statistical description for observed black holes that are generally rotating, non-extremal, neutral, non-supersymmetric, and asymptotically de Sitter (dS). 

The Bekenstein-Hawking area term can be considered as the tree-level term of the black hole entropy. Indeed, evaluating the Euclidean gravitational path integral with a classical solution led to the Bekenstein-Hawking area term without needing to know the details of microscopic configurations \cite{Gibbons1977}. This means that a microscopic picture of the black hole entropy must not only reproduce the Bekenstein-Hawking area term as the leading order but also lead to the corrections to it. It has been widely believed that the black hole entropy derived from the microscopic counting should have a general form as follows \cite{Chatterjee2020}
\begin{eqnarray}
 S_{\text{bh}}&=&\frac{A}{4G_N}\left(1+\cdots\right)+\alpha\ln A+\gamma\frac{4}{A}+\exp\{-\delta A\}+\cdots,\nonumber\\ \label{BHE-exp}
\end{eqnarray}
where $A$ refers to the area of the black hole event horizon, $G_N$ is Newton's gravitational constant, $\alpha$, $\gamma$, $\delta$, etc., are the universal constants, and the ellipses refer to the higher-order corrections. The logarithmic and exponential corrections have been found in \cite{Kaul2000,Jeon2017,Xiao2022} and \cite{Chatterjee2020,Murthy2015}, respectively. The ellipsis in the parentheses of Eq. (\ref{BHE-exp}) refers to the corrections of Newton's gravitational constant. Indeed, these corrections of the black hole entropy are unavoidable in calculating the statistical entropy of the black hole, but they have not been exhibited in this situation. The presence of these corrections can be easily understood as follows: Newton's gravitational constant is related to the gravitational energy of black holes which is the average energy of the thermodynamic system and is computed in the microstate picture as $\langle E\rangle=-\partial_\beta\log Z(\beta)$ where $Z(\beta)$ is a partition function corresponding to a statistical ensemble and $\beta$ is the inverse temperature; $\langle E\rangle$ includes both a leading term corresponding to the saddle-point approximation of the Euclidean gravitational path integral and the corrections beyond this approximation leading to the corrections of $G_N$.

It is expected that a microscopic description of the black hole entropy would be explained in an ultraviolet (UV) complete theory of quantum gravity. Unfortunately, such a theory of quantum gravity has not yet been available in a final form. However, when the spacetime curvature approaches the quantum gravity scales, the quantum fluctuations of the spacetime geometry play a crucial role in governing the behavior of physical objects. This means that quantum geometry effects could cause a transition from a black hole to a certain quantum one in which the unphysical singularity and the inconsistency of the black hole evaporation with the unitarity principle of quantum mechanics would be absent. On the other hand, the concept of black holes with the unphysical curvature singularity and the event horizon is absent in full quantum gravity, and it emerges only in the semiclassical limit \cite{Kiefer2020}. Indeed, some proposals imply the modification of the classical description of black holes within the microscopic distance of the horizon due to the presence of some new structure originating from quantum gravity, such as massive remnants \cite{Giddings1992}, non-local physics \cite{Giddings2006}, firewall \cite{Almheiri2013}, or fuzzball \cite{Kostas2008}.

For the above reasons, it would be interesting to explore the possibility of studying the microscopic configurations and statistical entropy of black holes in an intermediate regime between quantum gravity and general relativity without considering full quantum gravity. This work will show a novel microscopic description of the black hole entropy, which is understood in the framework of the compactified extra dimensions \cite{Antoniadis1990,Arkani-Hamed1998,Randall1999,Nam2021}. A derivation of the statistical entropy of four-dimensional (4D) black holes based on the compactification from higher to four dimensions is old. It has been found in the M-/string theory compactification to four dimensions \cite{Maldacena1996,Maldacena1997,Sen2008}. However, the microscopic configurations are M/D-branes that are BPS states, and hence the statistical entropy is only computed for (near-)extremal and supersymmetric black holes. In the present work, by considering the circle compactification of five-dimensional (5D) Einstein gravity with a positive cosmological constant, we will point to novel microscopic configurations that are not based on the context of M/D-branes. With this microscopic description, we calculate the statistical entropy of observed black holes that reproduces the Bekenstein-Hawking area term as the leading order and also includes the sub-leading corrections.

\section{Quantization of 5D gravity compactification}

In this section, we shall show that the compactification of the 5D Einstein-Hilbert (EH) action plus a positive cosmological constant on a circle $S^1$ leads to a discrete spectrum for the circle radius. As a result, there is a set of 4D gravitational configurations, each of which is basically characterized by a quantized value of the circle radius and represents a possible configuration that the observed 4D geometry might be in. This implies that 4D gravitational configurations arising from the dimensional reduction on the circle with the quantized radius spectrum would constitute a statistical ensemble of 4D geometries where we can use conventional statistical physics to study them. As a result, physical quantities observed in the four-dimensional world must be calculated by taking an average over the whole of this statistical ensemble. Here, the average values of the ensemble are calculable because of the discrete spectrum of 4D gravitational configurations. This insight will allow us to study the thermodynamics of 4D black holes in terms of statistical physics and calculate their statistical entropy, which is given in the next section.

Let us start with the gravitational action in five dimensions as follows
\begin{eqnarray}
S&=&\frac{M^3_*}{2}\int d^5X\sqrt{|g_5|}\left(\mathcal{R}^{(5)}-2\Lambda_5\right),\label{EHact}
\end{eqnarray}
where $M_*$ is the 5D Planck scale, $X^M$ are the 5D coordinates, $g_5$ is the determinant of the 5D metric, $\mathcal{R}^{(5)}$ stand for the 5D Ricci scalar, and $\Lambda_5$ refers to the bulk cosmological constant which is positive. By considering the circle compactification, the bulk metric can be generally decomposed as follows
\begin{eqnarray}
ds^2_5=g_{\mu\nu}dx^\mu dx^\nu-\phi^2\left[d\theta+g_{_A}A_\mu dx^\mu\right]^2,\label{KKmetric}
\end{eqnarray}
where $g_{\mu\nu}$, $A_\mu$, and $\phi$ are the 4D tensor, 4D vector, and 4D scalar components of the bulk metric, respectively, which are in general dependent on $(x^\mu,\theta$), and $g_{_A}$ is the gauge coupling. Using Eq. (\ref{KKmetric}), one can express $\mathcal{R}^{(5)}$ in terms of the 4D component fields (see detailed derivation in Ref. \cite{Nam2023,Nam2023b}) as follows
\begin{eqnarray}
\mathcal{R}^{(5)}&=&\hat{\mathcal{R}}+\frac{1}{4\phi^2}\left(\partial_\theta g^{\mu\nu}\partial_\theta g_{\mu\nu}+g^{\mu\nu}g^{\rho\lambda}\partial_\theta g_{\mu\nu}\partial_\theta g_{\rho\lambda}\right)-\frac{g^2_{_A}\phi^2}{4}F_{\mu\nu}F^{\mu\nu},\label{curexp}
\end{eqnarray}
where $\hat{\mathcal{R}}\equiv g^{\mu\nu}(\hat{\partial}_\lambda\hat{\Gamma}^\lambda_{\nu\mu}-\hat{\partial}_\nu\hat{\Gamma}^\lambda_{\lambda\mu}+\hat{\Gamma}^\rho_{\nu\mu}\hat{\Gamma}^\lambda_{\lambda\rho}-\hat{\Gamma}^\rho_{\lambda\mu}\hat{\Gamma}^\lambda_{\nu\rho})$ with $\hat{\Gamma}^\rho_{\mu\nu}\equiv\frac{g^{\rho\lambda}}{2}(\hat{\partial}_\mu g_{\lambda\nu}+\hat{\partial}_\nu g_{\lambda\mu}-\hat{\partial}_\lambda g_{\mu\nu})$, $\hat{\partial}_\mu\equiv\partial_\mu-g_{_A}A_\mu\partial_\theta$, and $F_{\mu\nu}=\partial_\mu A_\nu-\partial_\nu A_\mu$. Note that the kinetic term of the radion field $\phi$ is absent in the expansion of the 5D Ricci scalar $\mathcal{R}^{(5)}$ because its derivative terms are related to a total derivative. We can realize this from a geometric perspective: the circle has zero curvature, hence the derivative terms of the radion field should be absent in the expansion of the bulk curvature $\mathcal{R}^{(5)}$. However, the kinetic term of $\phi$ would appear when changing to the Einstein frame.

It is important to emphasize that the second term in Eq. (\ref{curexp}) is usually ignored in the literature because the dependence of the 4D components of the bulk metric on the fifth dimension is not considered. However, in this work, we will show that this term is essential to build microscopic configurations of observed black holes and compute their statistical entropy.

First, we determine the wavefunction profile of the 4D metric component, which describes its behavior along the fifth dimension. In order to do this, we consider the theory in the vacuum $\langle\phi\rangle=\text{constant}\equiv R$ that corresponds to $F_{\mu\nu}F^{\mu\nu}=0$ or $\langle A_\mu\rangle=0$ obtained from the equations of motion for $\phi$ and $A_\mu$ \cite{Overduin1997}. Then, the equations of motion for the 4D tensor component are found as follows \cite{Nam2023}
\begin{eqnarray}
\mathcal{R}-4\Lambda_5-\frac{1}{4R^2}\left[6g^{\mu\nu}\partial^2_\theta g_{\mu\nu}+4\partial_\theta g^{\mu\nu}\partial_\theta g_{\mu\nu}+5\left(g^{\mu\nu}\partial_\theta g_{\mu\nu}\right)^2\right]&=&0,\label{4Dtensor-equ}
\end{eqnarray}
where $\mathcal{R}\equiv g^{\mu\nu}(\partial_\lambda\Gamma^\lambda_{\nu\mu}-\partial_\nu\Gamma^\lambda_{\lambda\mu}+\Gamma^\rho_{\nu\mu}\Gamma^\lambda_{\lambda\rho}-\Gamma^\rho_{\lambda\mu}\Gamma^\lambda_{\nu\rho})$ with $\Gamma^\rho_{\mu\nu}\equiv\frac{g^{\rho\lambda}}{2}(\partial_\mu g_{\lambda\nu}+\partial_\nu g_{\lambda\mu}-\partial_\lambda g_{\mu\nu})$. The equation (\ref{4Dtensor-equ}) can be solved by the variable separation as $g_{\mu\nu}(x,\theta)=\chi(\theta)g^{(4)}_{\mu\nu}(x)$ where $g^{(4)}_{\mu\nu}(x)$ is identified as the metric of 4D spacetime and $\chi(\theta)$ is its wavefunction profile. This leads to
\begin{eqnarray}
\mathcal{R}^{(4)}&=&4\lambda,\label{effEinsEq}\\
3\chi''+8\frac{\chi'^2}{\chi}+2\kappa^2\chi&=&2\lambda R^2,\label{chiEq}
\end{eqnarray}
where $\mathcal{R}^{(4)}$ is the scalar curvature of 4D spacetime, $\kappa\equiv\sqrt{\Lambda_5}R$, and $\lambda$ is a constant. Eq. (\ref{effEinsEq}) means that the geometry of 4D spacetime is sourced by a cosmological constant $\lambda$ originating from the dynamics of the 4D tensor component of the bulk metric along the fifth dimension. The solution of Eq. (\ref{chiEq}) is given by
\begin{eqnarray}
 \chi(\theta)=\frac{11\lambda}{19\Lambda_5}\left[1-\cos\left(\sqrt{\frac{2}{11}}\kappa\theta\right)\right].\label{chi-solut}
\end{eqnarray}
The $S^1$ topology of the fifth dimension implies $\chi(\theta)=\chi(\theta+2\pi)$, which leads to the quantization for the size of the fifth dimension as follows
\begin{eqnarray}
R=\sqrt{\frac{11}{2\Lambda_5}}n \ \ \ \ \text{with} \ \ n=1,2,3,\cdots.\label{rad-quant}
\end{eqnarray}
The quantization rule (\ref{rad-quant}) means that the size of the fifth dimension is not arbitrary but must obtain discrete values. Recently, such a quantization relation has been found in the Swampland program \cite{Palti2019} and used to understand the radiative stability of the observed tiny cosmological constant \cite{Nam2023b}.

\section{\label{statphy}Statistical physics of 4D black holes}

\subsection{Statistical ensemble of 4D gravitational configurations}

The wavefunction profile of the 4D metric along the fifth dimension is characterized by two quantum numbers $n$ and $\lambda$. However, due to the non-linear nature of Eq. (\ref{chiEq}) originating from that of the metric, the solution of $g_{\mu\nu}(x,\theta)$ is not a linear combination of partial solutions. Hence, each value of the pair $n$ and $\lambda$ would lead to a possible gravitational configuration for the observed 4D geometry and it corresponds to a 4D effective action $S^{\{n,\lambda\}}_{\text{4D}}$ derived from  the dimensional reduction of Eq. (\ref{EHact}) on $S^1$ as follows
\begin{equation}
S^{\{n,\lambda\}}_{\text{4D}}=\frac{M^2_n}{2}\int d^4x\sqrt{-g_4}\left(\mathcal{R}^{(4)}-2\lambda\right)+\cdots,\label{n-4D-EHact}
\end{equation}
where $g_4$ is the determinant of the 4D metric $g^{(4)}_{\mu\nu}(x)$, $\mathcal{R}^{(4)}_{\mu\nu}$ is the Ricci tensor of 4D spacetime, the 4D Planck scale $M_n$ is given by
\begin{eqnarray}
M^2_n&=&M^3_*R\int^\pi_{-\pi}d\theta\chi(\theta)\nonumber\\
&=&\frac{4\pi\lambda M^3_*}{19}\left(\frac{11}{2\Lambda_5}\right)^{3/2}n,
\end{eqnarray}
and the ellipsis refers to the fluctuations of the radion and graviphoton fields around the vacuum $\langle\phi\rangle=R$ and $\langle A_\mu\rangle=0$. 

A set $\{S^{\{n,\lambda\}}_{\text{4D}}\}$ forms a statistical ensemble of the 4D gravitational configurations, whose average would lead to the observed 4D geometry. In this sense, the observed 4D geometry is realized as a macroscopic object that is described by quantities related to macroscopic or average properties such as the gravitational energy, the entropy, or the angular momentum. In this work, the macroscopic geometry which we are interested in is the 4D black holes which are neutral, rotating, and asymptotically dS.

As seen later, the quantum number $\lambda$ does not play a role in counting. It means that the statistical ensemble $\{S^{\{n,\lambda\}}_{\text{4D}}\}$ is basically characterized by the quantum number $n$. The interesting and novel point here is that due to the discrete spectrum of the size of the fifth dimension as indicated by Eq. (\ref{rad-quant}), this statistical ensemble is countable and the corresponding partition function is calculable in order to obtain a finite result. We would be unable to do this in the case that the spectrum of the size of the fifth dimension is continuous. This case corresponds to the limit of $\Lambda_5\rightarrow0$ where the solution (\ref{chi-solut}) becomes $\chi(\theta)=\lambda R^2\theta^2/19$ and thus lacks the $2\pi$ periodicity. Therefore, the positive bulk cosmological constant, $\Lambda_5>0$, is essential to provide a statistical physics description and compute the statistical entropy for the 4D black holes.

\subsection{Partition function}

The partition function $Z(\beta)$ of the statistical ensemble is computed by taking over all contributions from all possible 4D gravitational configurations that satisfy the specific boundary conditions: (i) fixing the ensemble temperature $T=1/\beta$ corresponding to the fact that the gravitational system is in equilibrium; (ii) fixing the event horizon radius $r_+$; (iii) fixing the asymptotic behavior; (iv) fixing the angular momentum if the gravitational system rotates. By using the Euclidean path integral formalism for the gravitational system \cite{Gibbons1977}, the partition function $Z(\beta)$ is evaluated on the Euclidean counterpart of the 4D Lorentzian geometry with the given boundary conditions that the Euclidean time $t_E=-it$ (with $t$ being the Lorentzian time) is periodic as $t_E\sim t_E+\beta$. Then, we obtain the partition function $Z(\beta)$ as follows
\begin{eqnarray}
Z(\beta)=\sum_{n=1}^{\infty}\int^{+\infty}_0 d\lambda\rho(\lambda)e^{-S^{\{n,\lambda\}}_E[g^{(4)}]},\label{Gen-partfunc}
\end{eqnarray}
where $\rho(\lambda)$ denotes the density of states corresponding to the continuous spectrum of $\lambda$ and $S^{\{n,\lambda\}}_E[g^{(4)}]$ is the Euclidean gravitational action (including the bulk and boundary terms and counterterms) which is related to the probability of finding the observed 4D geometry in the corresponding gravitational configuration determined by $\exp\{-S^{\{n,\lambda\}}_E[g^{(4)}]\}$.

In order to find the density of states $\rho(\lambda)$, we note the presence of a scale invariance in Eqs. (\ref{effEinsEq}) and (\ref{chiEq}) as follows
\begin{eqnarray}
g^{(4)}_{\mu\nu}\rightarrow bg^{(4)}_{\mu\nu},\ \ \chi\rightarrow\chi/b,\ \ \lambda\rightarrow\lambda/b,\label{scale-trans}
\end{eqnarray}
where $b$ is a scale parameter. Accordingly, the partition function $Z(\beta)$ given by Eq. (\ref{Gen-partfunc}) is also invariant under this scale transformation. This suggests that the density of states is given by the Dirac delta function as $\rho(\lambda)=\delta(\lambda)$, as a result of the scale invariance \cite{Jensen2011}. It means that an observed 4D cosmological constant $\langle\lambda\rangle$, which is calculated by the average of the statistical ensemble, reads
\begin{eqnarray}
\langle\lambda\rangle=\frac{1}{Z(\beta)}\sum_{n=1}^{\infty}\int^{+\infty}_0 d\lambda\rho(\lambda)\lambda e^{-S^{\{n,\lambda\}}_E[g^{(4)}]}=0.   
\end{eqnarray}
However, when taking into account the presence of the matter perturbations, the scale invariance (\ref{scale-trans}) would no longer be exact, but it is an approximate symmetry. On the other hand, the observed 4D cosmological constant $\langle\lambda\rangle$ should be near-zero, instead of vanishing. This leads to a small shift in the Dirac delta function of the density of states as 
\begin{eqnarray}
\rho(\lambda)=\delta(\lambda-\lambda_0),\label{density}    
\end{eqnarray}
where $\lambda_0=\langle\lambda\rangle\neq0$.

With the quantized spectrum of the size of the fifth dimension and the density of states $\rho(\lambda)$ given by Eqs. (\ref{rad-quant}) and (\ref{density}), respectively, the partition function $Z(\beta)$ given by the expression (\ref{Gen-partfunc}) is computable precisely and is derived as follows
\begin{eqnarray}
Z(\beta)=\frac{1}{e^{\tilde{S}_E}-1},\label{Part-Func-one}
\end{eqnarray}
where $\tilde{S}_E\equiv S^{\{1,\lambda_0\}}_E[g^{(4)}]$. 

\subsection{Thermodynamic quantities}

The free energy $F$ of the 4D black hole is determined by the partition function $Z(\beta)$ as
\begin{eqnarray}
F&=&-\frac{1}{\beta}\log Z(\beta)\nonumber\\
&=&\frac{1}{\beta}\left[\tilde{S}_E+\log\left(1-e^{-\tilde{S}_E}\right)\right].
\end{eqnarray}
Then, from the standard thermodynamics as $dF=-TdS$ with $S$ being the black hole entropy, one can obtain the entropy as
\begin{eqnarray}
S&=&-\partial_TF=\left(1-\beta\partial_\beta\right)\log Z(\beta)\nonumber\\
&\simeq&-\partial_T\left(\tilde{S}_ET\right)\left(1+e^{-\tilde{S}_E}\right)+\left(1+\tilde{S}_E\right)e^{-\tilde{S}_E},\label{bh-stat-entro}
\end{eqnarray}
where in the second line we have expanded in $e^{-\tilde{S}_E}$, which is much smaller than one with respect to observed 4D black holes. The first term $-\partial_T\left(\tilde{S}_ET\right)$ corresponds to the saddle-point approximation of the gravitational path integral \cite{Gibbons1977}, which would reproduce the Bekenstein-Hawking area term at leading order. The remaining terms would lead to the sub-leading corrections, which include the exponential corrections.

Besides the exponential corrections found above, the black hole entropy can obtain the logarithmic corrections. An interesting additional contribution to the partition function comes from the quantum fluctuations around classical solutions \cite{Hawking1978}. Evaluating the functional integral that is quadratic in the quantum fluctuations around classical solutions by using the heat kernel leads to the logarithmic correction for the black hole entropy \cite{Fursaev1996,Sen2013,El-Menoufi2016}. The logarithmic correction also appears at the first order when considering the influence of the generalized uncertainty principle, which indicates a minimal length scale below which the spacetime geometry is distorted \cite{RAli2024,Babar2024,RAli2025,Babar2025}. 

The energy of the 4D black hole, which corresponds to the average energy of the statistical ensemble, is
\begin{eqnarray}
\langle E\rangle &=&F+TS=-\partial_\beta\log Z(\beta)\nonumber\\
&\simeq&\partial_\beta\tilde{S}_E\left(1+e^{-\tilde{S}_E}\right).\label{ave-energ}
\end{eqnarray}
The first term $\partial_\beta\tilde{S}_E$ is the classical contribution for the energy of the 4D black hole, corresponding to the saddle-point approximation, which is familiar in the literature. The second term $\sim e^{-\tilde{S}_E}$ is a new contribution which describes the exponential correction for the energy of the 4D black hole.

\section{The statistical entropy of 4D Kerr-$\text{dS}$ black holes}

Let us do explicit calculations for the statistical entropy of 4D Kerr-\text{dS} black holes which are neutral, rotating, and live in the asymptotically dS spacetime. The black hole solution was found by Carter \cite{Carter1968} and is known as a special case of the Plebanski-Demianski family of metrics \cite{Plebanski1976}. The 4D Kerr-dS black hole geometry is characterized by three macroscopic quantities that are the gravitational energy, the rotational parameter $a$, and the asymptotically dS radius $l=\sqrt{3/\lambda_0}$. For an asymptotically flat extremal Kerr black hole, the Bekeinstein-Hawking entropy without the corrections could be produced from the microscopic counting in string theory by mapping this black hole into a nonrotating Kaluza-Klein black hole \cite{Horowitz2007}.

Now, we will use the results obtained in Sec. \ref{statphy} to compute the statistical entropy of the 4D Kerr-dS black holes that are generally non-extremal and non-supersymmetric. First, we need to calculate the Euclidean gravitational action associated with each 4D gravitational configuration as follows
\begin{eqnarray}
S^{\{n,\lambda_0\}}_E[g^{(4)}]=I^{(n)}_{\text{bulk}}+I^{(n)}_{\text{surf}}+I^{(n)}_{\text{ct}},
\end{eqnarray}
where $I^{(n)}_{\text{bulk}}$ is the Euclidean EH action derived by the Wick rotation of the action (\ref{n-4D-EHact}), $I^{(n)}_{\text{surf}}$ is the Gibbons-Hawking-York term \cite{Gibbons1977,York1972} given by
\begin{eqnarray}
I^{(n)}_{\text{surf}} =-M^2_n\int_{\partial M}d^3x\sqrt{h}K,   
\end{eqnarray}
where $\partial M$, $h$, and $K$ refer to the boundary of 4D spacetime, the determinant of the induced metric on $\partial M$, and the trace of extrinsic curvature of $\partial M$, respectively, and $I^{(n)}_{\text{ct}}$ is the counterterm whose contribution would produce a finite Euclidean gravitational action and is given by the extension of the AdS counterterm in Ref. \cite{Emparan1999} for the dS case as
\begin{eqnarray}
I^{(n)}_{\text{ct}}=-M^2_n\int_{\partial M}d^3x\sqrt{h}\left[\frac{2}{l}-\frac{l}{2}\mathcal{R}_3+\frac{l^3}{2}\left(\frac{3}{8}\mathcal{R}^2_3-\mathcal{R}_{3ab}\mathcal{R}^{ab}_3\right)\right],
\end{eqnarray}
where $\mathcal{R}_3$ and $\mathcal{R}_{3ab}$ are the Ricci scalar and Ricci tensor of $\partial M$, respectively. With the 4D Kerr-dS black hole metric written in Boyer-Lindquist type coordinates \cite{Akcay2011}, we find
\begin{eqnarray}
S^{\{n,\lambda_0\}}_E[g^{(4)}]=\frac{2\pi M^2_n\beta}{l^2\Xi}\left(r^3_++\Xi l^2r_++\frac{l^2a^2}{r_+}\right),\label{kerr-ds-Eact}
\end{eqnarray}
where $\Xi=1+a^2/l^2$ and the inverse temperature $\beta$ is given as
\begin{eqnarray}
\beta=\frac{4\pi(r^2_++a^2)}{r_+}\left(1-\frac{a^2}{l^2}-\frac{a^2}{r^2_+}-3\frac{r^2_+}{l^2}\right)^{-1}.
\end{eqnarray} 

By using Eq. (\ref{bh-stat-entro}) with $\tilde{S}_E=S^{\{1,\lambda_0\}}_E[g^{(4)}]$ derived in Eq. (\ref{kerr-ds-Eact}), the statistical entropy of the 4D Kerr-dS black hole is found as follows
\begin{eqnarray}
S&=&\frac{A}{4G_N}\left[1+\left(1+\frac{A}{\pi l^2}+\frac{8\pi a^2}{A}\right)e^{-\frac{A}{4G_N}\left(1+\frac{A}{\pi l^2}+\frac{8\pi a^2}{A}\right)}\right]\nonumber\\
&&+e^{-\frac{A}{4G_N}\left(1+\frac{A}{\pi l^2}+\frac{8\pi a^2}{A}\right)}+\cdots,\label{BHentr}
\end{eqnarray}
where $A=4\pi(r^2_++a^2)/\Xi$ is the horizon area of the 4D Kerr-dS black hole, the observed Newton's gravitational constant $G_N$ is given by $G_N\equiv1/(8\pi\langle M^2_n\rangle)$ with the average value $\langle M^2_n\rangle$ calculated as follows
\begin{eqnarray}
\langle M^2_n\rangle&=&\frac{1}{Z(\beta)}\sum_{n=1}^{\infty}\int^{+\infty}_0 d\lambda\rho(\lambda)M^2_ne^{-S^{\{n,\lambda\}}_E[g^{(4)}]}\nonumber\\
&=&M^2_1\left[1+e^{-\tilde{S}_E}+\mathcal{O}\left(e^{-2\tilde{S}_E}\right)\right],
\end{eqnarray}
and the ellipsis refers to the higher-order terms in the powers of $e^{-\tilde{S}_E}$, $A/l^2$, and $a^2/A$. We observe that the Bekenstein-Hawking area term is reproduced at leading order, and other terms are the sub-leading corrections due to the average of the statistical ensemble. The microscopic description of the 4D black hole in the present work predicts the exponential corrections, which are easy to see by setting $l^2\rightarrow\infty$ and $a=0$, corresponding to the entropy of the 4D Schwarzschild black hole, which is asymptotically flat as
\begin{eqnarray}
S=\frac{A}{4G_N}\left[1+e^{-\frac{A}{4G_N}}\right]+e^{-\frac{A}{4G_N}}+\cdots.\label{SBH-entr}    
\end{eqnarray}
The exponential correction corresponding to the last term has been found in the quantum representation of the horizon geometry, like in loop quantum gravity \cite{Chatterjee2020}, and it is also exhibited in string theory \cite{Murthy2015}. However, we here discover a new exponential correction which is given by the second term in square brackets and is more meaningful than the exponential correction found previously. From Eq. (\ref{BHentr}) or (\ref{SBH-entr}), one can realize that the new exponential correction arises due to the correction of Newton's gravitational constant that corresponds to the correction of the gravitational energy of the 4D black hole. This can be realized by observing the gravitational energy of the 4D black hole given by Eq. (\ref{ave-energ}) where the second term yields the new exponential correction of the black hole entropy.

\section{Summary and future directions}
In this work, we construct microscopic configurations that form a statistical ensemble for observed 4D black holes and compute their statistical entropy. By considering the circle compactification of Einstein gravity with a positive cosmological constant in five dimensions, we find a statistical ensemble of 4D gravitational configurations that is mainly classified in terms of the size of the fifth dimension. This result is not new in the literature. But the interesting and novel point is here that we demonstrate that the size of the fifth dimension must, in fact, be quantized. This means that the size of the fifth dimension is not arbitrary but must obtain discrete values according to the quantization rule. As a result, this statistical ensemble is countable, and hence we can compute precisely the gravitational partition function from which the thermodynamic quantities would be derived. This cannot be performed in the situation of the continuous spectrum with respect to the size of the fifth dimension. In particular, the statistical derivation of the black hole entropy in this work goes beyond special black holes which are supersymmetric and (near-)extremal as well as have exotic charges. Also, the present framework can be applied to compute the statistical entropy of general 4D black holes independent of the symmetries of the black hole solution, such as spherical symmetry.

The present approach, based on the quantization of the size of the fifth dimension, has implications for future studies. First, the emergence of the scale invariance (\ref{scale-trans}), which is exact in the case of the vacuum geometry and becomes approximate when including the matter perturbations, can provide a solution for the cosmological constant problem \cite{Weinberg1989,Bousso2008}. Second, with the statistical ensemble for the observed 4D black holes, one can calculate the statistical fluctuations around the average, which is identified as the observed 4D world. We can find the influence of these statistical fluctuations in the spectrum of the gravitational waves or the geodesic motion of particles around the observed 4D black holes. Third, if the equations of motion have more than one solution, then the partition function $Z(\beta)$ given by Eq. (\ref{Part-Func-one}) becomes
\begin{eqnarray}
Z(\beta)=\sum_i\frac{1}{e^{\tilde{S}^{(i)}_E}-1},\label{MulMG-partfunc}
\end{eqnarray}
where the index $i$ refers to the macroscopic geometries. Interestingly, there are phase transitions between these macroscopic geometries as the temperature is below a critical value, such as the Hawking-Page transition \cite{DPage1983} interpreted as the confinement/deconfinement phase transition in the AdS/CFT correspondence \cite{Witten1998}. Such phase transitions have been studied by using the semiclassical approximation, where the detail of the statistical physics is ignored. However, the partition function (\ref{MulMG-partfunc}) can provide a framework for understanding how microscopic configurations govern such phase transitions. It would be interesting to see what features of microscopic configurations might be revealable in these phase transitions.

\section*{Acknowledgments}
We would like to thank the referees for their valuable comments, suggestions, and questions, by which the quality of the paper has been improved. We appreciate JINR (Dubna) for the hospitality where part of this work was done.


\begin{thebibliography}{99}

\bibitem{Bekenstein1973} J. D. Bekenstein, Phys. Rev. D {\bf 7}, 2333 (1973). 

\bibitem{Bardeen1973} J. M. Bardeen, B. Carter, and S. W. Hawking, Commun. Math. Phys. {\bf 31}, 161 (1973).

\bibitem{Hawking1975} S. W. Hawking, Commun. Math. Phys. {\bf 43}, 199 (1975) [Erratum-ibid. {\bf 46}, 206 (1976)].

\bibitem{Strominger1996} A. Strominger and C. Vafa, Phys. Lett. B {\bf 379}, 99 (1996).

\bibitem{Horowitz1996} G. T. Horowitz and A. Strominger, Phys. Rev. Lett. {\bf 77}, 2368 (1996).

\bibitem{Maldacena1996} J. M. Maldacena and A. Strominger, Phys. Rev. Lett. {\bf 77}, 428 (1996).

\bibitem{Johnson1996} C. V. Johnson, R. R. Khuri, and R. C. Myers, Phys. Lett. B {\bf 378}, 78 (1996).

\bibitem{Dabholkar2005} A. Dabholkar, Phys. Rev. Lett. {\bf 94}, 241301 (2005).

\bibitem{Boido2023} A. Boido, J. P. Gauntlett, D. Martelli, and J. Sparks, Phys. Rev. Lett. {\bf 130}, 091603 (2023).

\bibitem{Ashtekar1998} A. Ashtekar, J. Baez, A. Corichi, and K. Krasnov, Phys. Rev. Lett. {\bf 80}, 904 (1998).

\bibitem{Ibrah2021} I. Bah and P. Heidmann, Phys. Rev. Lett. {\bf 126}, 151101 (2021).

\bibitem{Balasubramanian2023} V. Balasubramanian, A. Lawrence, J. M. Magan, and M. Sasieta, arXiv:2212.08623.

\bibitem{Cadoni2021} M. Cadoni, M. Oi, and A. P. Sanna, Phys. Rev. D {\bf 104}, L121502 (2021).

\bibitem{Gibbons1977}  G. W. Gibbons and S. W. Hawking, Phys. Rev. D {\bf 15}, 2752 (1977).

\bibitem{Chatterjee2020} A. Chatterjee and A. Ghosh, Phys. Rev. Lett. {\bf 125}, 041302 (2020).

\bibitem{Kaul2000} R. K. Kaul and P. Majumdar, Phys. Rev. Lett. {\bf 84}, 5255 (2000).

\bibitem{Jeon2017} I. Jeon and S. Lal, Phys. Lett. B {\bf 774}, 41 (2017).

\bibitem{Xiao2022} Y. Xiao and Y. Tian, Phys. Rev. D {\bf 105}, 044013 (2022).

\bibitem{Murthy2015} A. Dabholkar, J. Gomes, and S. Murthy, JHEP {\bf 1503} (2015) 074.

\bibitem{Kiefer2020} C. Kiefer, J. Phys.: Conf. Ser. {\bf 1612}, 012017 (2020).

\bibitem{Giddings1992} S. B. Giddings, Phys. Rev. D {\bf 46}, 1347 (1992). 

\bibitem{Giddings2006} S. B. Giddings, Phys. Rev. D {\bf 74}, 106005 (2006).

\bibitem{Almheiri2013} A. Almheiri, D. Marolf, J. Polchinski, J. Sully, JHEP {\bf 02} (2013) 062.

\bibitem{Kostas2008} K. Skenderis and M. Taylor, Phys. Rept. {\bf 467}, 117 (2008).

\bibitem{Antoniadis1990}  I. Antoniadis, Phys. Lett. B {\bf 246}, 377 (1990).

\bibitem{Arkani-Hamed1998} N. Arkani-Hamed, S. Dimopoulos, and G. R. Dvali, Phys. Lett. B {\bf 429}, 263 (1998).

\bibitem{Randall1999} L. Randall and R. Sundrum, Phys. Rev. Lett. {\bf 83}, 3370 (1999).

\bibitem{Nam2021} C. H. Nam, Eur. Phys. J. C {\bf 81}, 1102 (2021).

\bibitem{Maldacena1997} J. Maldacena, A. Strominger, and E. Witten, JHEP {\bf 9712} (1997) 002.

\bibitem{Sen2008} A. Sen, Gen. Rel. Grav. {\bf 40}, 2249 (2008).

\bibitem{Nam2023} C. H. Nam, Phys. Rev. D {\bf 107}, L041901 (2023).

\bibitem{Nam2023b} C. H. Nam,  Phys. Rev. D {\bf 107}, 063502 (2023).

\bibitem{Overduin1997} J. M. Overduin and P. S. Wesson, Phys. Rept. {\bf 283}, 303 (1997).

\bibitem{Palti2019} E. Palti, Fortschr. Phys. {\bf 67}, 1900037 (2019).

\bibitem{Jensen2011} K. Jensen, S. Kachru, A. Karch, J. Polchinski, and E. Silverstein, Phys. Rev. D {\bf 84}, 126002 (2011).

\bibitem{Hawking1978} S. W. Hawking, Phys. Rev. D {\bf 18}, 1747 (1978).

\bibitem{Fursaev1996} D. V. Fursaev and S. N. Solodukhin, Phys. Lett. {\bf B365}, 51 (1996).

\bibitem{Sen2013} A. Sen, JHEP {\bf 04} (2013) 156.

\bibitem{El-Menoufi2016} B. K. El-Menoufi, JHEP {\bf 05} (2016) 035.


\bibitem{RAli2024} R. Ali, X. Tiecheng, and R. Babar, Nucl. Phys. B {\bf 1008}, 116710 (2024). 

\bibitem{Babar2024} R. Ali, X. Tiecheng, and R. Babar, Gen. Rel. Grav. {\bf 56}, 13 (2024).

\bibitem{RAli2025} R. Ali, X. Tiecheng, and R. Babar, Phys. Dark Univ. {\bf 48}, 101868 (2025). 

\bibitem{Babar2025} R. Ali, X. Tiecheng, and R. Babar, Commun. Theor. Phys. {\bf 77}, 075404 (2025).

\bibitem{Carter1968} B. Carter, Commun. Math. Phys. {\bf 10} (1968) 280.

\bibitem{Plebanski1976} J. F. Plebanski and M. Demianski, Annals Phys. {\bf 98}, 98 (1976).


\bibitem{Horowitz2007} G. T. Horowitz and M. M. Roberts, Phys. Rev. Lett. {\bf 99}, 221601 (2007).

\bibitem{York1972} J. W. York, Phys. Rev. Lett. {\bf 28}, 1082 (1972).

\bibitem{Emparan1999} R. Emparan, C. V. Johnson, and R. C. Myers, Phys. Rev. D {\bf 60}, 104001 (1999). 

\bibitem{Akcay2011} S. Akcay and R. A. Matzner, Class. Quant. Grav. {\bf 28}, 085012 (2011).

\bibitem{Weinberg1989} S. Weinberg, Rev. Mod. Phys. {\bf 61}, 1 (1989).

\bibitem{Bousso2008} R. Bousso, Gen. Relativ. Gravit. {\bf 40}, 607 (2008).

\bibitem{DPage1983} S. W. Hawking and D. N. Page, Comm. Math. Phys. {\bf 87}, 577 (1983).

\bibitem{Witten1998} E. Witten, Adv. Theor. Math. Phys. {\bf 2} (1998) 505.


\end{thebibliography}
\end{document}